\documentclass[runningheads,citeauthoryear]{apinv38}
\usepackage{epsfig,cite,graphics}
\usepackage{marvosym} 
\usepackage[utf8]{inputenc}

\begin{document}

\title{The hidden symbiotic star SU Lyn - detection of flickering in U band}
\titlerunning{Flickering of the symbiotic-like binary star SU Lyn}
\author{R. K. Zamanov\inst{1}, 
        A. Kostov\inst{1},
        M. Moyseev\inst{1}, 
 	N. Petrov\inst{1}, 
  	Y. M. Nikolov\inst{1},
	G. Y. Latev\inst{1}, 
        D. Marchev\inst{2}, 
	S. Boeva\inst{1},  
 	K. A. Stoyanov\inst{1}, 
        M. S. Minev\inst{1,3},  
        J. Mart\'i\inst{4},	
        V. Radeva\inst{5}, 
	E. S\'anchez-Ayaso\inst{6}, 
        M. F. Bode\inst{7,8}, 
	K. I{\l}kiewicz\inst{9},   
	G. Nikolov\inst{1}, 
	P. L. Luque-Escamilla\inst{10},
	B. Spassov\inst{1},
	B. Borisov\inst{2},  
        V. D. Marchev\inst{1}, 
        A. Kurtenkov\inst{1}
	\vskip 0.3cm 
	$ \; $
	}
\authorrunning{Zamanov et al.}
\tocauthor{R. K. Zamanov, A. Kostov, M. Moyseev, N. Petrov, 
Y. M. Nikolov, D. Marchev, K. A. Stoyanov, M. S. Minev, V. Radeva, B. Borisov,  
K. I{\l}kiewicz, M. F. Bode, G. Nikolov, A. Kurtenkov, G. Y. Latev, S. Boeva, P. L. Luque-Escamilla} 
\institute{Institute of Astronomy and National Astronomical Observatory, 
Bulgarian Academy of Sciences, Tsarigradsko Shose 72, BG-1784 Sofia, Bulgaria 
 \and 
 Department of Physics and Astronomy, Shumen University "Episkop Konstantin Preslavski",  
  115 Universitetska Str., BG-9700 Shumen, Bulgaria
 \and 
 Department of Astronomy,  Faculty of Physics, Sofia University "Saint Kliment Ohridski",  
 5 James Bourchier Blvd., BG-1164 Sofia, Bulgaria
 \and 
  Departamento de F{\'i}sica (EPSJ), Universidad de Ja{\'e}n, Campus Las Lagunillas s/n, A3, E-23071 Ja{\'e}n, Spain  
 \and 
 Naval Academy, 73 Vasil Drumev str., BG-9026 Varna, Bulgaria
 \and 
 Departamento de Ciencias Integradas, Centro de Estudios Avanzados en F{\'i}sica, Matem{\'a}tica y Computaci{\'o}n, 
 Universidad de Huelva, E-21071, Huelva, Spain
 \and 
 Astrophysics Research Institute, Liverpool John Moores University, 
 IC2, 149 Brownlow Hill, Liverpool, L3 5RF, UK
 \and 
  Office of the Vice Chancellor, Botswana International University of Science and Technology, 
  Private Bag 16, Palapye, Botswana
 \and  
 Astronomical Observatory, University of Warsaw, Al. Ujazdowskie 4, 00-478 Warszawa, Poland
 \and 
  Departamento de Ingenier{\'i}a Mec{\'a}nica y Minera (EPSJ), 
  Universidad de Ja{\'e}n, Campus Las Lagunillas s/n, A3, E-23071 Ja{\'e}n, Spain
  \vskip 0.3cm 
  $ \; $
  }
\papertype{Submitted on 21 June 2022; Accepted on 11 August 2022}	
\maketitle

\begin{abstract}
We report photometric observations of the hidden symbiotic star SU~Lyn in the optical bands.  
In five nights we detect a weak  flickering in U band with amplitude of about 0.05 magnitudes.
No intranight variations are found in B, V, g' and r' bands.  
This is one more indication that the secondary component is a white dwarf accreting at a low accretion rate. 

We also searched for intranight variability of a dozen related objects
- RR Boo, RT Boo, AM Cyg, AG Peg, BF Cyg, NQ Gem, StHa190, V627~Cas, XX~Oph, FS~Cet and Y Gem 
- however no variability above the observational errors is detected.   
\end{abstract}
\vskip 0.32cm 

\section{Introduction}
The symbiotic stars are interacting binaries with long
orbital periods in the range from 100 days to more than 100 years. 
They consist of a red or yellow star transferring mass to a hot compact object
(e.g. Miko{\l}ajewska 2012). 
The mass donor is a  giant or supergiant of spectral class G-K-M.  
If the giant is an Asymptotic Giant Branch star, the system usually is a strong infrared source. 
More than 250 symbiotic systems are known in our Galaxy (Akras et al. 2019, Merc et al. 2019).

Flickering is a non-periodic variability (on minutes-to-hour timescales)
typical for the accreting white dwarfs in  cataclysmic variables (e.g. Bruch 1992, 2021). 
Systematic searches for flickering in symbiotic binaries (Dobrzycka et al. 1996;
Sokoloski et al. 2001) showed that 
flickering  is rarely detected in the classical symbiotic stars. 
Until now, only 11 of the symbiotic stars 
are observed to vary in the optical bands and do so on a timescale
of  $\sim 10$ minutes with amplitude  $> 0.1$ magnitude -- 
RS~Oph, T~CrB, MWC~560, Z~And, V2116~Oph, CH~Cyg,
RT~Cru, o~Cet, V407~Cyg, V648~Car, EF~Aql, and ZZ~CMi. 
The last three were identified as flickering sources during the last decade: 
V648~Car (Angeloni et al. 2012),  EF~Aql (Zamanov et al. 2017), 
and ZZ~CMi (Zamanov et al. 2021). 

Here we report 
(i) photometry of SU~Lyn and detection of intranight varibility in U band
and
(ii) photometry of 11 related objects in which no variability is detected.

\begin{table*}
\caption{Photometric observations of SU Lyn. In the table are given date
(in format YYYY-MM-DD), telescope, band, UT start-end,
number of the frames and exposure time, 
 typical observational error (merr),  type of variability.
}             
\centering
\begin{tabular}{lcccc c  c | c cc cc} 
\hline
&          &             &    &               &                & $ \; $       &	 $\;$  & &			     & \\
  date   &  telescope  & band &  UT start-end &   frames       & $ \; $ mag   & merr $\;$  & &   variability	     &  \\   
&          &             &    &               &                & $ \; $       &	 $\;$  & &			     & \\
\hline       
&          &             &    &               &                & $ \; $       &	 $\;$  & &			     & \\
2020-01-18 & 50/70 Roz   &  U & 23:53 - 02:23 & 173 x 30 sec   & $ \; $ 10.24 & 0.007$\;$  & &  flickering 0.11 mag      & \\
2020-04-07 & 2.0m        &  U & 18:21 - 20:24 &  60 x 120 sec  & $ \; $ 10.90 & 0.002$\;$  & &  flickering 0.05 mag      & \\
2020-04-08 & 2.0m        &  U & 18:04 - 20:38 & 260 x 30 sec   & $ \; $ 10.87 & 0.004$\;$  & &  flickering 0.05 mag      & \\
2020-04-17 & 2.0m        &  U & 18:12 - 20:28 & 170 x 45 sec   & $ \; $ 11.22 & 0.020$\;$  & &  flickering 0.03 mag      & \\
           &             &  V & 18:14 - 20:27 & 648 x 0.02 sec & $ \; $  8.15 & 0.005$\;$  & &  no variability           & \\
2021-03-13 & 50/70 Roz   &  U & 23:42 - 01:41 &  98 x 30 sec   & $ \; $ 10.24 & 0.008$\;$  & &  smooth     0.2 mag       & \\  
           &             &  B & 23:43 - 01:42 &  97 x 10 sec   & $ \; $  9.96 & 0.005$\;$  & &  no variability           & \\ 
2022-03-27 & 60cm        &  U & 20:02 - 22:42 & 273 x 15 sec   & $ \; $ 11.22 & 0.015$\;$  & &  flickering 0.03 mag      & \\
           &             &  B & 20:02 - 22:42 & 273 x  2 sec   & $ \; $ 10.05 & 0.007$\;$  & &  no variability  	 & \\
2022-04-27 & 40cm Shu    &  g'& 18:31 - 21:15 & 282 x 10 sec   & $ \; $ 9.26  & 0.010$\;$  & &  no variability           & \\
           &             &  r'& 18:30 - 21:15 & 295 x  4 sec   & $ \; $ 7.68  & 0.010$\;$  & &  no variability           & \\
&          &             &    &               &                & $ \; $       &	     $\;$  & &			         & \\
\hline                                                           
  \end{tabular}                                                  
  \label{tab1}
\end{table*}

 \begin{figure*}[]     
  \vspace{5.9cm}   
  \includegraphics{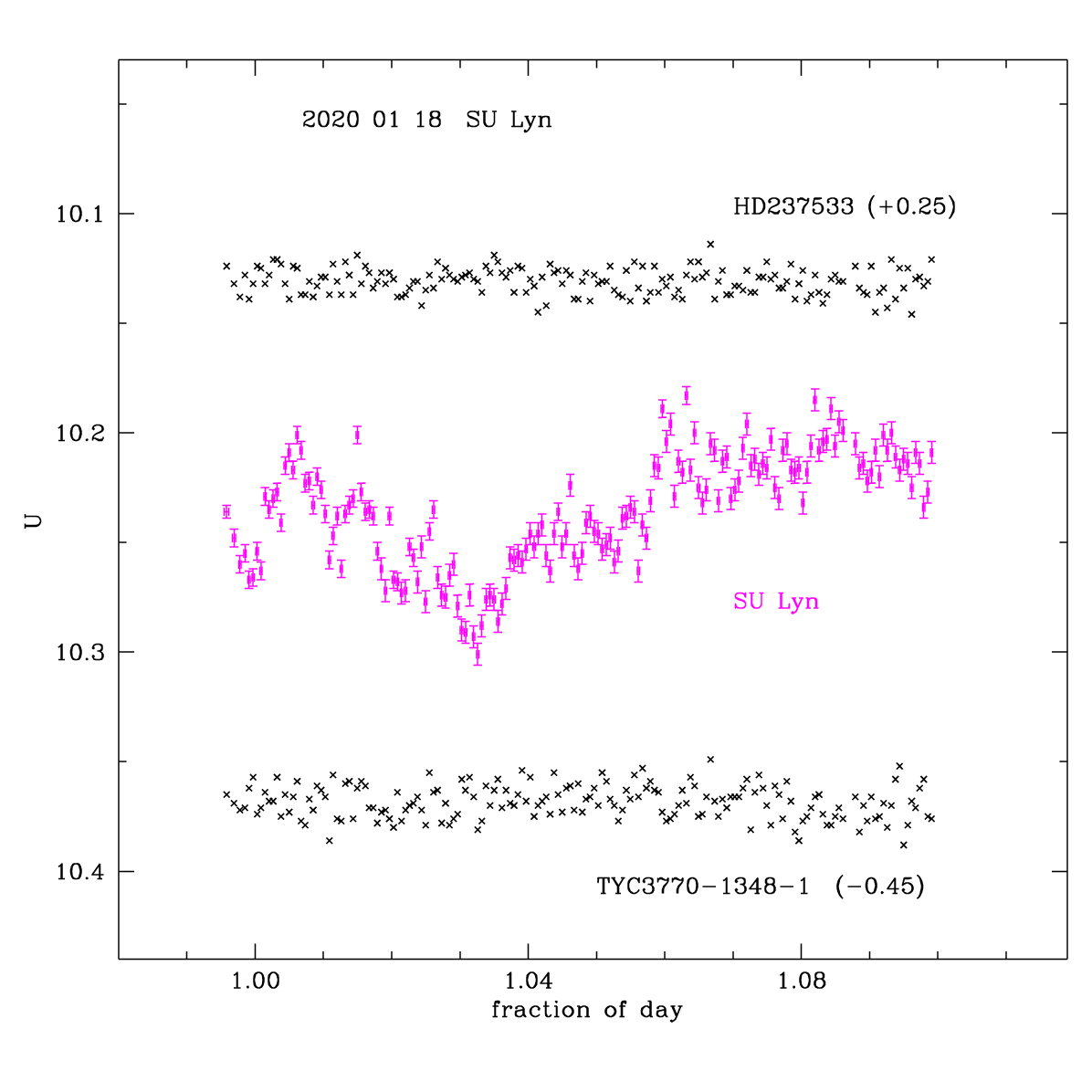}     
  \includegraphics{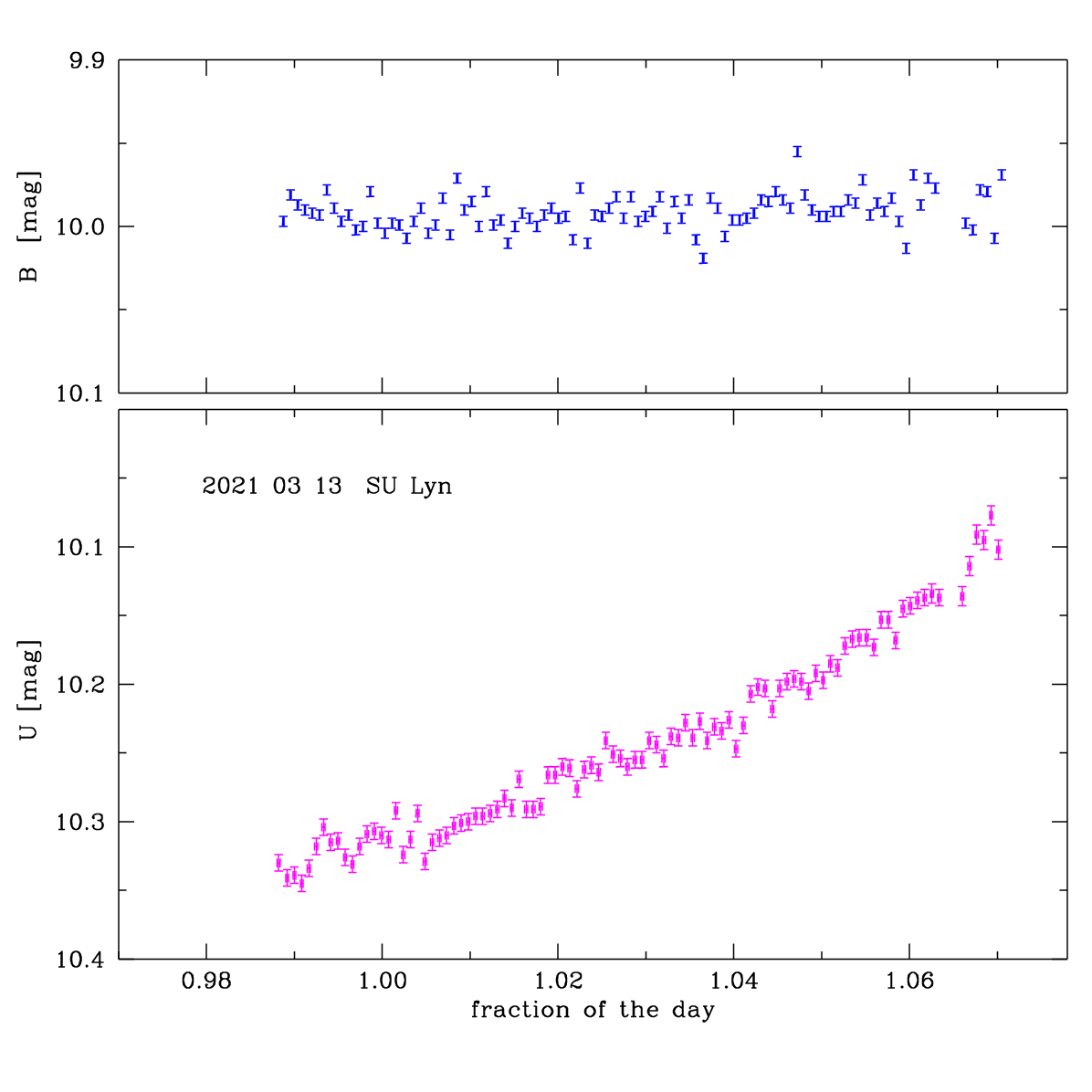}   
  \caption[]{Detection of intranight variability of SU Lyn with the 50/70 cm Schmidt telescope 
  of NAO Rozhen in U band. }
\label{fig1}      
  \vspace{3.5cm}   
  \includegraphics{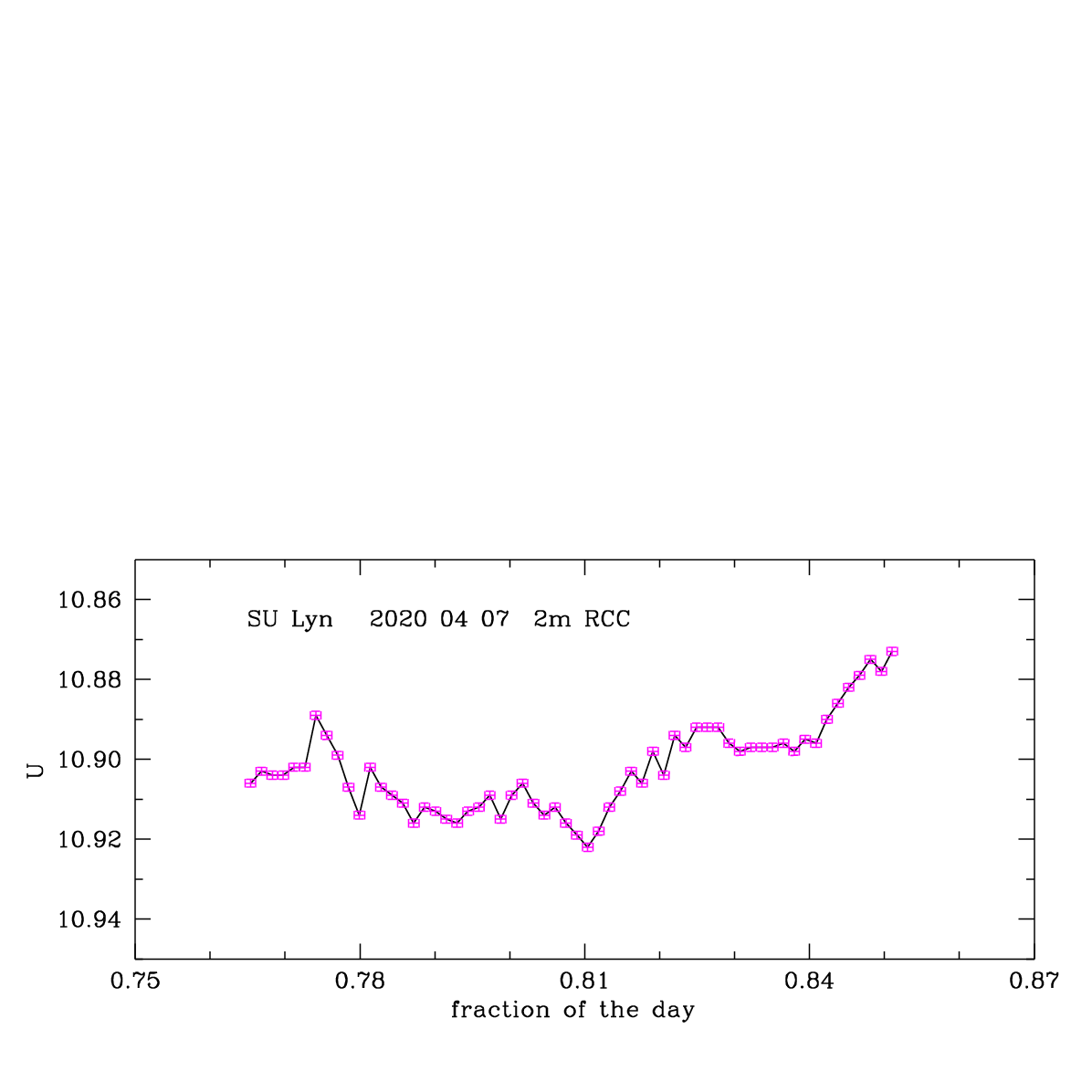}     
  \includegraphics{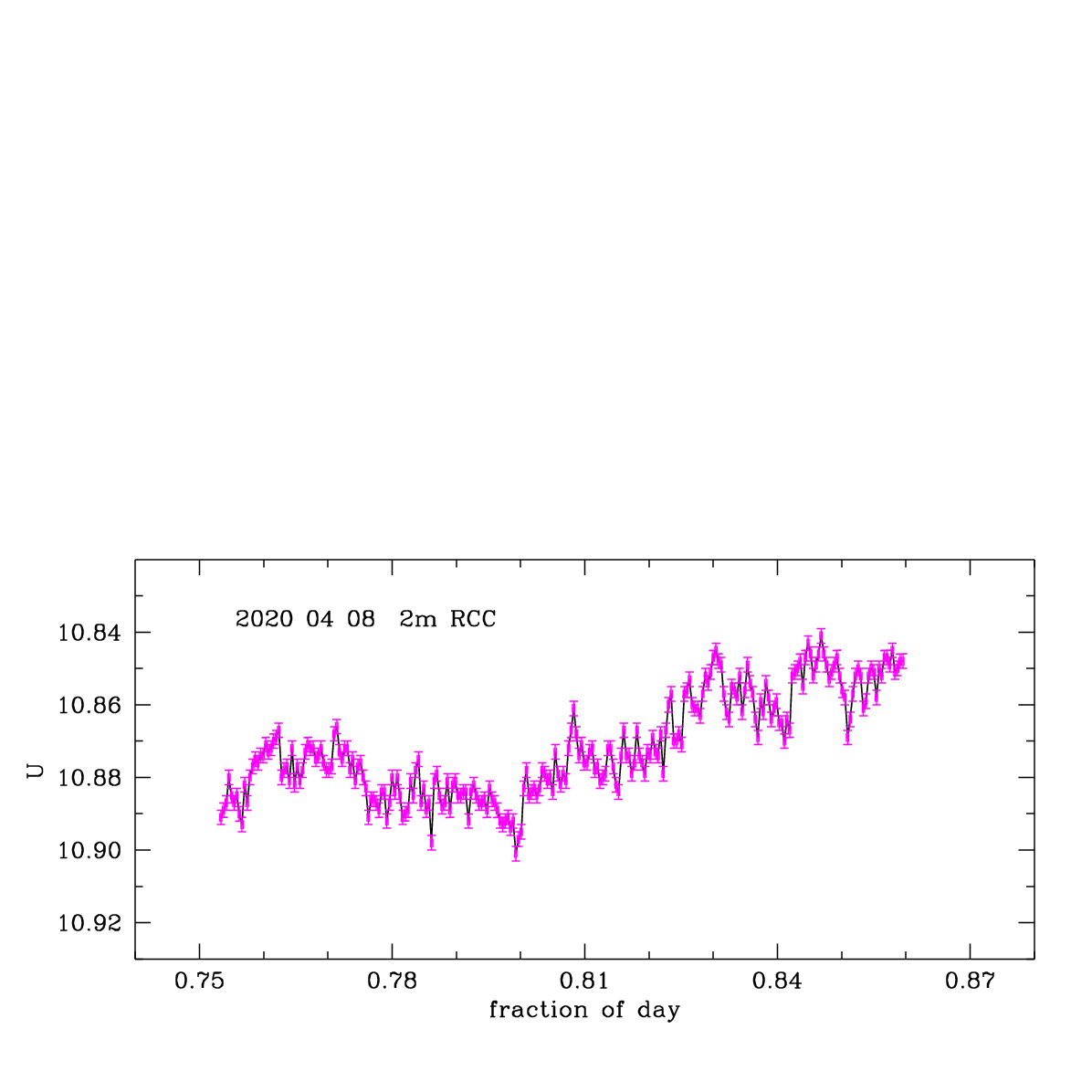}   
  \caption[]{Low amplitude flickering with $\Delta U \approx 0.05$ is visible in the observations 
   with the  2.0m telescope in April 2020.   }
\label{fig2}      
  \vspace{6.2cm}   
  \includegraphics{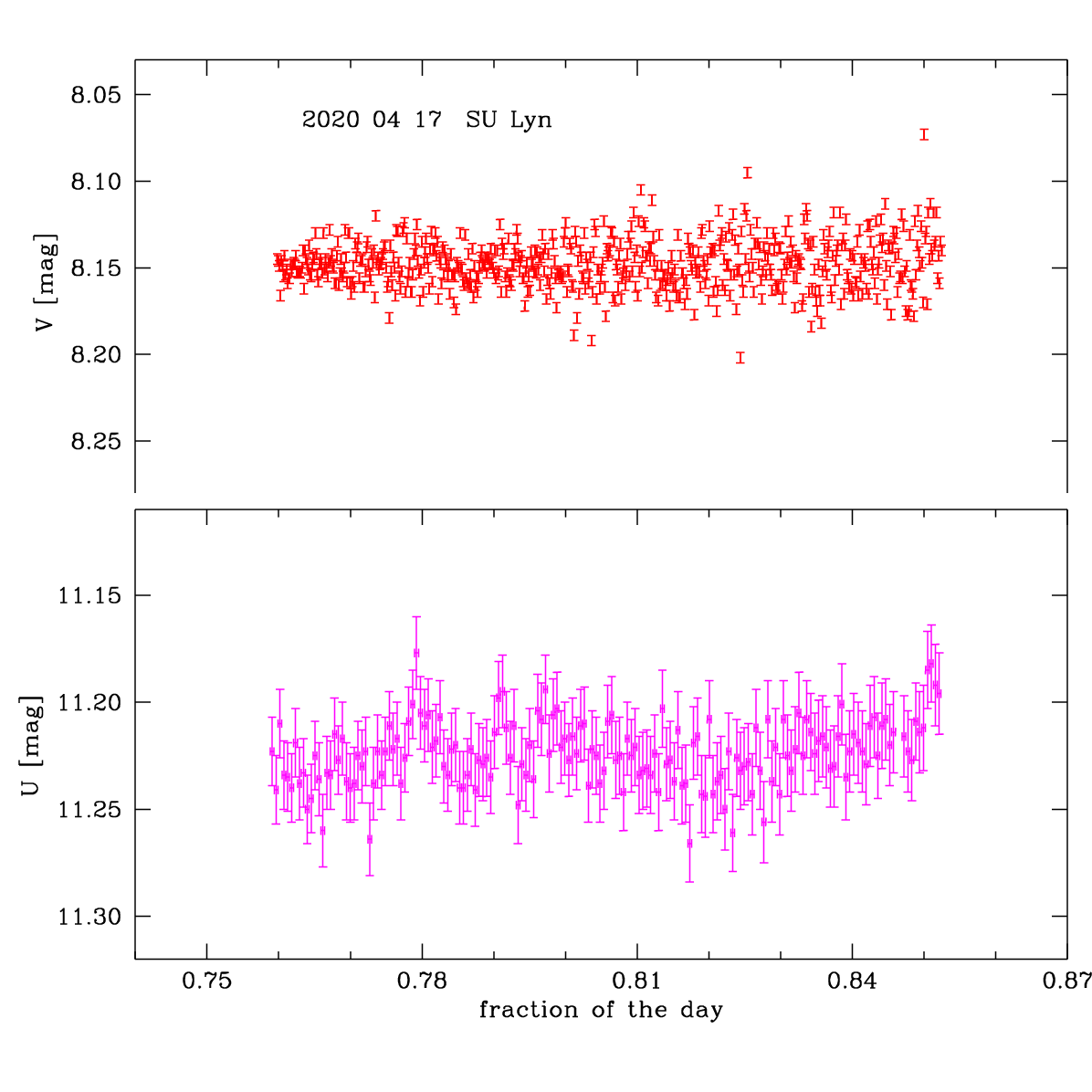}     
  \includegraphics{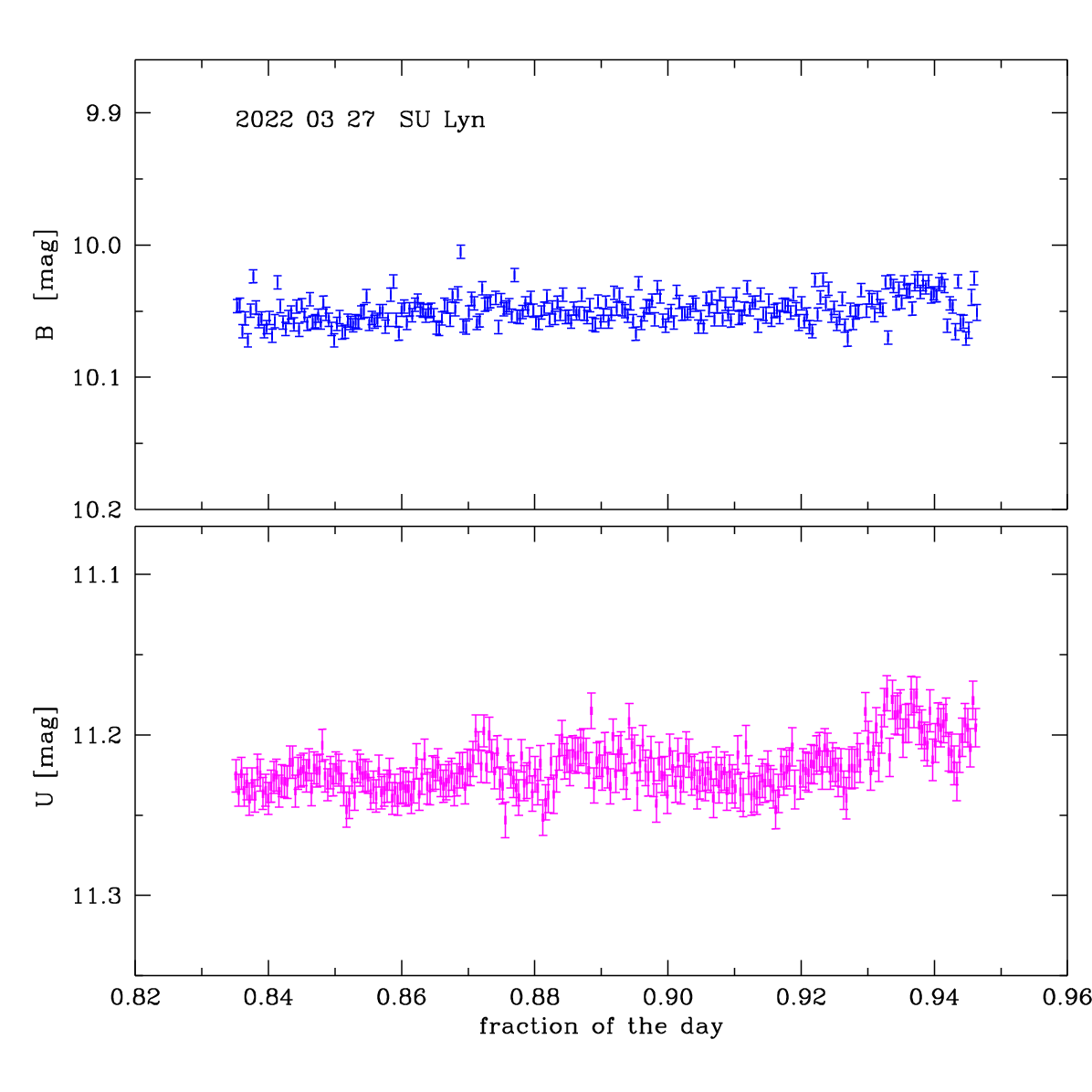}   
  \caption[]{Simultaneous observations in two bands. 
  A low amplitude  variability is visible in U band, with amplitude $\Delta U \approx 0.03$~mag.
  No variability is detected in B and V bands.  }
\label{fig3}      
\end{figure*}	     

\begin{figure}   
  \vspace{7.0cm}   
  \includegraphics{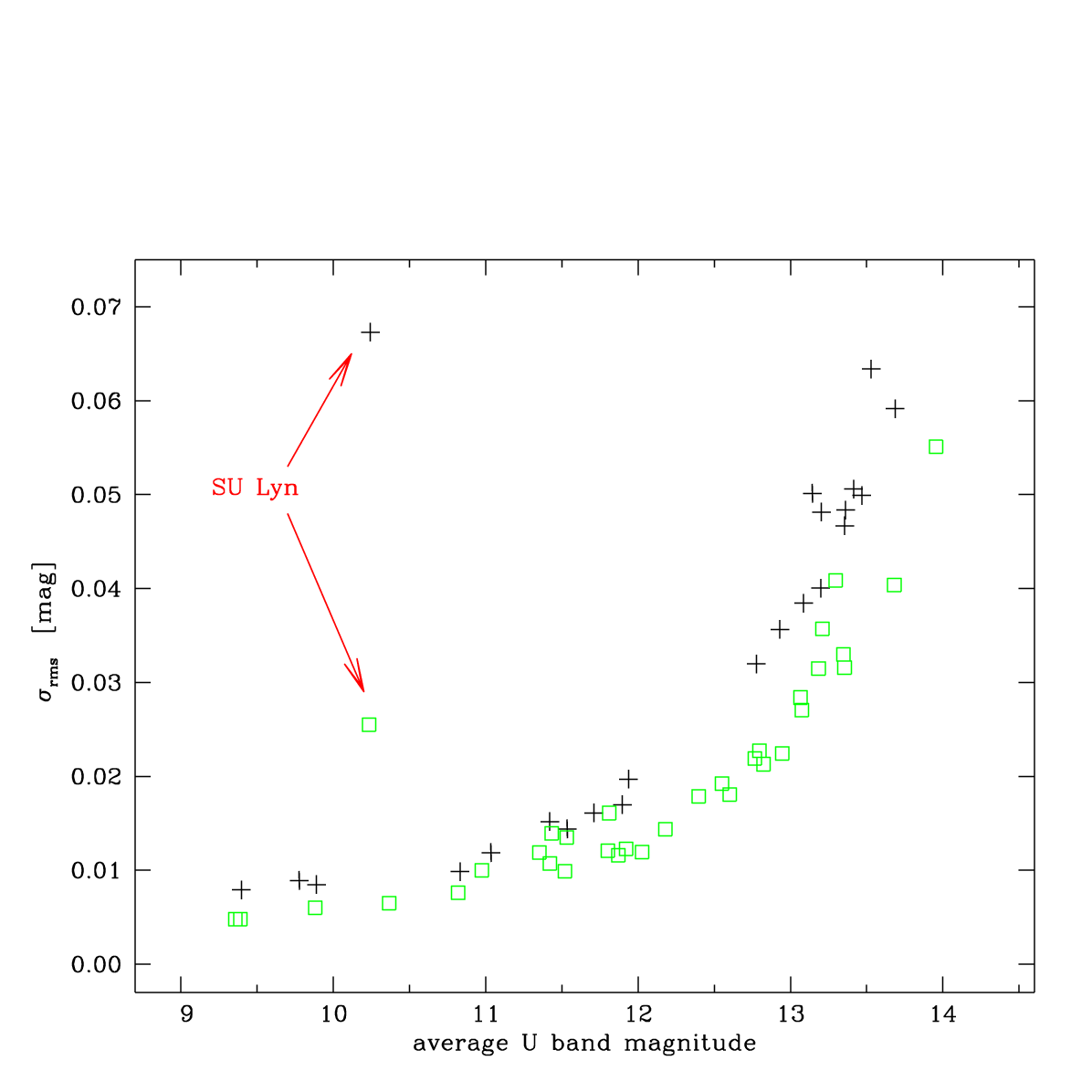} 
  \caption[]{Root mean square deviation versus the average U-band magnitude. 
             The green squares refer to the night 2020-01-18, 
	     and the plus signs refer to 2021-03-13.     
	     The rms of SU Lyn deviates from the behavior
             of the field stars, which indicates that it is variable. 
	     }
\label{fig.rms}      
\end{figure}  

\begin{table*}
\caption{Log of observations on which no variability is detected. 
In the Table are given object, date of observation, telescope, band, UT start - UT end, 
number of the frames and exposure time, average magnitude, 
typical observational error (merr),  and the result (no variability).
}             
\centering
\begin{tabular}{lcc | cccc | crc |  c cc cc} 
\hline

object   &  date-obs   &  &  & telescope & band &  UT start-end  &  &  frames	   &  &  mag    &  merr  &  result  & \\
         &             &  &  &		 &	&		 &  &		   &  &	        &        & 	    & \\
         &             &  &  &		 &	&		 &  &		   &  &	        &        & 	    & \\
RR Boo   & 2019-06-21  &  &  & 41cm Jaen &  B	&  21:10 - 22:01 &  &  67 x 20 sec &  &  11.99  &  0.01  & negative & \\
         &             &  &  &		 &  V	&		 &  &  66 x 10 sec &  &  10.21  &  0.01  & negative & \\
         &             &  &  &		 &	&		 &  &		   &  &	        &        & 	    & \\
RT Boo   & 2019-06-22  &  &  & 41cm Jaen &  B	&  20:59 - 21:35 &  &  34 x 30 sec &  &  14.03  &  0.03  & negative & \\
         &             &  &  &		 &  V	&		 &  &  32 x 20 sec &  &  12.09  &  0.01  & negative & \\
         &             &  &  &		 &	&		 &  &		   &  &	        &        & 	    & \\
AM Cyg   & 2019-08-27  &  &  & 2.0m Roz	 &  V   &  23:04 - 23:59 &  & 281 x 10 sec &  &  14.29  &  0.01  & negative & \\
         &             &  &  &		 &   	&		 &  &              &  &	        &        &          & \\
\hline
         &             &  &  &		 &   	&		 &  &              &  &	        &        &          & \\       
	 &             &  &  &		 &   	&		 &  &              &  &	        &        &          & \\
AG Peg   & 2017-09-26  &  &  & 41cm Jaen &  B	&  18:44 - 19:51 &  & 145 x 10 sec &  &   9.71  &  0.01  & negative & \\
         &             &  &  &		 &  V	&		 &  & 137 x 5  sec &  &   8.56  &  0.01  & negative & \\
         &             &  &  &		 &	&		 &  &		   &  &	        &        & 	    & \\
BF Cyg   & 2017-09-18  &  &  & 41cm Jaen &  B	&  18:59 - 20:06 &  &  87 x 20 sec &  &  10.11  &  0.01  & negative & \\
         &             &  &  &		 &  V	&		 &  &  87 x 20 sec &  &   9.66  &  0.01  & negative & \\
         &             &  &  &		 &	&		 &  &		   &  &	        &        & 	    & \\
NQ Gem   & 2018-03-22  &  &  & 41cm Jaen &  B	&  20:04 - 21:01 &  &  79 x 15 sec &  &  10.16  &  0.01  & negative & \\
         &             &  &  &		 &  V	&  20:33 - 21:01 &  &  52 x 5 sec  &  &   8.05  &  0.01  & negative & \\
         &             &  &  &		 &	&		 &  &		   &  &	        &        & 	    & \\
StHa190  & 2017-10-02  &  &  & 41cm Jaen &  B	&  19:02 - 20:07 &  &  53 x 30 sec &  &  11.37  &  0.01  & negative & \\
         &             &  &  &		 &  V	&		 &  &  53 x 30 sec &  &  10:54  &  0.01  & negative & \\ 
	 &             &  &  &		 &	&		 &  &		   &  &	        &        & 	    & \\
V627 Cas & 2017-09-18  &  &  & 50/70 Roz &  B   & 01:03 - 02:34  &  & 160 x 30 sec &  &  14.75  &  0.03  & negative & \\       
         &             &  &  &		 &   	&		 &  &              &  &	        &        &          & \\
\hline
         &             &  &  &		 &	&		 &  &		   &  &	        &        & 	    & \\
Y Gem    & 2018-10-09  &  &  & 60 cm Bel &  B	&  02:17 - 03:21 &  &  71 x 30 sec &  &  10.90  &  0.01  & negative & \\
         &             &  &  &		 &  V	&		 &  &  71 x 5 sec  &  &   9.51  &  0.01  & negative & \\
         &             &  &  &		 &   	&		 &  &              &  &	        &        &          & \\
FS Cet   & 2015-08-17  &  &  & 50/70 Roz &  B   & 00:54 - 01:47  &  & 130 x 20 sec &  &  12.44  &  0.01  & negative & \\ 
         & 2016-12-19  &  &  & 50/70 Roz &  V   & 18:05 - 19:01  &  & 240 x 10 sec &  &  12.35  &  0.01  & negative & \\
	 &             &  &  &		 &   	&		 &  &              &  &	        &        &          & \\ 
XX Oph   & 2018-09-14  &  &  & 50/70 Roz &  U   & 17:36 - 18:15  &  &  39 x 30 sec &  &  10.25  &  0.01  & negative & \\    
         &             &  &  & 50/70 Roz &  B   & 17:38 - 18:12  &  &  35 x 5 sec  &  &   9.87  &  0.01  & negative & \\
	 &             &  &  &		 &	&		 &  &		   &  &	        &        & 	    & \\  
\hline                                                           
  \end{tabular}                                                  
  \label{tab2}
\end{table*}

\section{Observations}
\label{s.obs}

We have observed SU Lyn with 
(i)   the 50/70 cm Schmidt telescope  of the National Astronomical Observatory (NAO) Rozhen, Bulgaria, 
(ii)  the 2.0 m RCC telescope of the NAO Rozhen,  
(iii) the 60 cm telescope of the  NAO Rozhen,  
      and 
(iv)  the 40 cm telescope of the Shumen University  ``Episkop Konstantin Preslavski", Bulgaria 
      (Kjurkchieva et al. 2020).  
All the telescopes are equipped with CCD cameras.

The observations are obtained in broad bands in Johnson and SDSS systems:   
U band --   effective central wavelength 4671 \AA, 
B band --  4380 \AA, 
V band --   5445 \AA, 
g' band -- 4671 \AA,  
r' band -- 6141 \AA. 
More details about the filter transmission can be found in  Rodrigo, Solano \& Bayo (2012).    
As comparison stars we used HD47726 (U=9.36$^*$, B=8.69, V=7.72)
and HD47954 (U=8.90, B=8.90, V=8.41),
HD237533 (U=9.88$^*$, B=9.78, V=9.20).  
The magnitudes are taken from SIMBAD, 
except those marked with $^*$, which are our estimations using HD47954.  

%
%

The observations are plotted on Fig.~\ref{fig1}, Fig.~\ref{fig2}, and Fig.~\ref{fig3}.
It has to be noted that on the left panel of Fig.~\ref{fig1} are plotted two other stars --
HD~237533  and  TYC~3770-1348-1. 
HD~237533 has an average  U=9.88 and is brighter than SU~Lyn.
TYC~3770-1348-1 has an average  U=10.82  and  is fainter than SU~Lyn. 
Both are shifted to appear on the same scale as SU~Lyn. 

Table~\ref{tab1} is a journal of observations of SU~Lyn, where are given
the date when the observations begin, the telescope, the filter, 
UT-start and UT-end of the run, number of the frames and exposure time,
typical observational error (in magnitudes), and the type of variability
visible in the run.

We also observed another 11 objects 
-- three variable stars of Mira-type (RR Boo, RT Boo and AM~Cyg),  a sample
of symbiotic stars (AG Peg, BF Cyg, NQ Gem, StHa190, V627~Cas) as well as three other objects (Y~Gem, FS~Cet,  XX~Oph). 
These observations are performed also with: \\
-- (1)  the telescope of the University of Ja\'en (Spain). This is an astronomical facility located within the University of Ja\'en 
premises and mostly devoted to educational purposes. It hosts an automated 41 cm telescope
equipped with a CCD camera with Johnson-Cousins filters. Differential photometry in the
B and V bands was conducted as described in Mart\'{\i} et al. (2017).  
\\
-- (2) the 60 cm telescope of the Belogradchik Astronomical Observatory, Bulgaria
(Strigachev \& Bachev 2011).  
The log of observations with non-detections is contained in Table~\ref{tab2}.

\vskip 0.3cm

\section{Intranight variability of SU~Lyn in U band}

\label{s.res}

Variability on a time scale of 1 hour is detected in all U band observations. 
In 4 nights, it is a low amplitude flickering, with amplitude in the range $0.03 < \Delta U \le 0.11$~mag.
In one night we observe a smooth increase of the brightness (2021-03-13, right panel of Fig.~\ref{fig1})
in U band of 0.2 mag over 2 hours, while no variability is visible 
in the simultaneous observations in B band. 

The intranight variability is detected only in U band.
No variability on a time scale of 1 hour is visible on our runs 
at longer wavelengths (B, V, g', r'). 
It is likely that the red giant is the dominating source in these bands.

For the two runs obtained with the Rozhen Schmidt telescope (presented on Fig.~\ref{fig1})
we measure the standard deviation of SU Lyn and of about 20
other stars in the field and plot it on Fig.~\ref{fig.rms}. 
These two runs are appropriate because the field of view of the 
Schmidt telescope is wide and and a sufficient number of stars can be measured.  
The standard deviation is calculated as 
\begin{equation}
    \sigma_{rms} = \sqrt{ \frac{1}{N_{pts}-1} \sum\limits_{i}(m_i - \overline m )^2 } ,
    \label{eq.sig}
\end{equation}
where $\overline m$ is the average magnitude in the run, $N_{pts}$ is the number of the data points.
The $\sigma_{rms}$ calculated in this way includes 
the variability of the star (if it exists) 
and the measurement errors. 
For non-variable stars it represents  the precision of the photometry.
The rms deviation of SU~Lyn  is 5-7 times larger
than that expected from observational errors, 
indicating that SU~Lyn is variable on a timescale of 
$\sim 1$ hour. 


SU Lyn is a binary system composed of a white dwarf and a red giant. 
Lopes de Oliveira et al. (2018) estimate that the mass of the white dwarf 
is $\ge 0.7$~M$_\odot$.
In the optical wavelengths there are no 
prominent emission lines typical for symbiotic stars -- H$\alpha$ is visible as an weak emission, with equivalent width of about  $-0.5$~\AA.  
SU Lyn is bright and variable at X-ray wavelengths, with 
periods of high X-ray luminosity and strong UV variability
and classified  as a hidden symbiotic star (Mukai et al. 2016).
I{\l}kiewicz et al. (2022) proposed  that 
SU Lyn can be a progenitor of a classical, persistent symbiotic system.

The presence of flickering in U band is one more piece of evidence that the hot component is a  white dwarf
surrounded by an accretion disc (see Sect. 4.1  in  Sokoloski \& Bildsten 2010).
For a comparison, the amplitudes of the flickering in U band of a few symbiotic stars are: 
 RS Oph  0.3 - 0.4 mag, 
 CH~Cyg  0.4 - 0.5 mag, 
 ZZ~CMi  0.05 - 0.25 mag.
SU~Lyn has a lower amplitude, which means that the accretion disc is not bright
and therefore the white dwarf is accreting at a lower accretion rate.

\vskip 0.3cm

\section{Search for intranight variability of related objects}

We also searched for intranight variability of a few related objects --
RR Boo, RT Boo, AM Cyg, AG Peg, BF Cyg, NQ Gem, StHa190, V627 Cas, FS Cet, XX~Oph and Y Gem.

Rapid brightness variations of the Miras  RR Boo, RT Boo and AM~Cyg are detected in Hipparcos photometry
with amplitude $\approx 0.3$~mag (de Laverny et al. 1998). 
However we do not detect a similar variability in our runs. 


AG Peg made a transition from a slow symbiotic nova (which drove the 1850 outburst) 
to a classical symbiotic star (Ramsay et al. 2016).  
The UV emission of AG Peg shows stochastic variability (flickering) 
on timescales of minutes and hours (Zhekov \& Tomov 2018). 
We do not detect variability in B and V bands.   

BF Cyg is a symbiotic star which displays signatures of a variable mass-outflow and 
formation of a highly-collimated bipolar mass ejection (Shchurova et al. 2019; Tomov et al. 2019). 
No variability is detected in B and V bands. 

NQ Gem displays  optical spectra similar to that of the symbiotic recurrent nova T CrB 
(Greene \& Wing 1971) and X-ray emission similar to that of CH Cyg (Luna et al. 2013). 
Both T~CrB and CH~Cyg are known flickering sources, however 
we do not detect similar variability of NQ Gem in our B and V band observations.

StHa190 is the first yellow symbiotic star with rapidly variable bipolar mass outflow 
(Munari et al. 2001). We do not detect variability on a timescale of 1 hour in B and V bands.   

V627 Cas is a symbiotic star (Vra{\v{s}}{\v{t}}{\'a}k  2018). We do not detect variability 
on a timescale of 1 hour in B band.

For Y Gem,  Sahai et al. (2018) find strong and variable UV and X-ray emission and strong flickering in the UV continuum. Snaid et al. (2018) observed flickering in u' and g' bands. 
We do not detect variability on a timescale of 1 hour in B and V bands. 

FS Cet is included in the Catalog of cataclysmic variables 
selected from the Large Sky Area Multi-Object Fiber Spectroscopic Telescope (LAMOST) Data Release 5
(Hou et al. 2020). Although the flickering is typical for cataclysmic varaibles, 
we do not detect variability on time scale of 1 hour in B and V bands.

XX Oph is known as Merrill's iron star (Howell et al. 2009). 
We do not detect variability on a timescale of 1 hour in U and B bands. 


\vskip 0.3cm

{\bf Conclusions: }
We report photometric observations of the symbiotic-like star SU~Lyn performed with
the 2.0~m and the 50/70~cm Schmidt telescopes of NAO Rozhen. 
In five nights we detect a low amplitude flickering in U band (amplitude in the range 0.05 - 0.10 mag).  
This indicates that the secondary component is a white dwarf accreting at low accretion rate. 

For a few related objects 
(RR Boo, RT Boo, AM Cyg, AG Peg, BF Cyg, NQ Gem, StHa190, V627 Cas, Y Gem, FS Cet, and XX Oph), 
the search for rapid optical variability (in the most cases in B and V bands) gives a negative result.


\vskip 0.3cm 

{\small {\bf Acknowledgements: }
This work is part of the project KP-06-H28/2 08.12.2018  "Binary stars with compact object"  
(Bulgarian National Science Fund).  
DM and BB acknowledge project RD-08-100/2022.  
JM and PLE acknowledge support from Programa Operativo FEDER 2014-2020 and
Consejer\'{\i}a de Econom\'{\i}a y Conocimiento of Junta de Andaluc\'{\i}a (Ref. 1380270).}


\newpage 

\end{document}